\documentclass[aps,prd,twocolumn,superscriptaddress,showpacs,floatfix,preprintnumbers,amsmath,amssymb]{revtex4}
\usepackage{epsfig}
\usepackage{color}
\input minisym

\begin{document}

\preprint{\babar-PUB-07/073}
\preprint{SLAC-PUB-13160}

\title{Measurement of the Spin of the $\Xi(1530)$ Resonance}

%
\author{B.~Aubert}
\author{M.~Bona}
\author{Y.~Karyotakis}
\author{J.~P.~Lees}
\author{V.~Poireau}
\author{X.~Prudent}
\author{V.~Tisserand}
\author{A.~Zghiche}
\affiliation{Laboratoire de Physique des Particules, IN2P3/CNRS et Universit\'e de Savoie, F-74941 Annecy-Le-Vieux, France }
\author{J.~Garra~Tico}
\author{E.~Grauges}
\affiliation{Universitat de Barcelona, Facultat de Fisica, Departament ECM, E-08028 Barcelona, Spain }
% ...
\collaboration{The \babar\ Collaboration}
\noaffiliation

\date{\today}

\begin{abstract}
The properties of the $\Xi(1530)$ resonance are investigated in the $\Lambda_c^+\rightarrow \Xi^- \pi^+ K^+$ decay process.
The data sample was collected with the \babar\ detector at
 the SLAC PEP-II asymmetric-energy $e^+ e^-$ collider operating at center
 of mass energies 10.58 and 10.54 GeV. The corresponding integrated
 luminosity is approximately 230 fb$^{-1}$.
The spin of the $\Xi(1530)$ is established to be 3/2.  The existence of an
$S$-wave amplitude in the $\Xi^- \pi^+$ system is inferred, and its interference with the
$\Xi(1530)^0$ amplitude provides the first clear demonstration of the Breit-Wigner phase motion
expected for the $\Xi(1530)$.  The $P_1(\rm cos\,\theta_{\Xi^-})$ Legendre polynomial 
moment indicates the presence of a significant $S$-wave amplitude for $\Xi^- \pi^+$ mass values above 1.6 GeV/c$^2$, 
and a dip in the mass distribution at approximately 1.7 GeV/c$^2$ is interpreted as due to coherent addition 
of a $\Xi(1690)^0$ contribution to this amplitude.  This would imply $J^P=1/2^-$ for the $\Xi(1690)$.  
Attempts at fitting the $\Xi(1530)^0$ lineshape yield unsatisfactory results, and this failure is attributed to interference effects 
associated with the amplitudes describing the $K^+ \pi^+$ and/or $\Xi^- K^+$ systems.
\end{abstract}

\maketitle

\section{INTRODUCTION}

The $\Xi(1530)$ is the only cascade resonance whose properties are reasonably well understood.
It decays $\sim$ 100\% to $\Xi \pi$ and $<$ 4\% to $\Xi \gamma$~\cite{PDG2006}, and its
mass (PDG fit: $m(\Xi(1530)^0)=1531.80\pm 0.32$ MeV/c$^2$) and width (PDG fit: $\Gamma(\Xi(1530)^0)=0.1\pm 0.5$ MeV/c) 
are reasonably well known~\cite{PDG2006}.
A spin-parity analysis of data on the reactions $K^- p \rightarrow \Xi(1530)^{0,-} K^{0,+}$
carried out by Schlein et {\it al.}~\cite{ref:schlein}
 showed that $J^P =3/2^+$ (i.e., $P$-wave) or $J^P=5/2^-$ (i.e., $D$-wave) is favored, and that the data are
 consistent with $J\ge 3/2$;
however, they stated that spin $>3/2$ is not required, and on this basis concluded that $J^P=3/2^+$.
This conclusion was supported by Button-Schafer et {\it al.}~\cite{ref:button}
in a similar analysis.  
Both experiments ruled out $J=1/2$, but the claim that $J>3/2$ was not required
was the basis for the conclusion that $J^P=3/2^+$.
In the present paper, the $\Omega^-$ spin analysis procedures described in Ref.~\cite{ref:omesp} 
are extended to the quasi-two-body decay $\Lambda_c^+ \rightarrow (\Xi^- \pi^{+}) K^{+}$,
for which the $\Xi^- \pi^{+}$ invariant mass distribution exhibits a dominant $\Xi(1530)^0$ signal~\cite{cc}.  
Under the assumption that the $\Lambda_c^+$ has spin 1/2, it is established that the $\Xi(1530)$ has spin 3/2.  
On the basis of the analyses of Refs.~\cite{ref:schlein,ref:button}, 
it follows that positive parity is established.     

The data sample and event selection procedures are described
in Section II, and the $\Xi(1530)$ spin measurement is
presented in Section III. In Section IV, the amplitude
structure in the $\Xi(1530)$ region is investigated in
some detail, and this is followed by an examination of the
$\Xi^- \pi^+$ system at higher mass values in Section V. The
unsuccessful attempts at precise measurements of the mass
and width of the $\Xi(1530)^0$ are presented in Section VI,
and their implications considered. Finally, the conclusions
drawn from this analysis are summarized in Section VII.

\section{THE \babar\ DETECTOR AND $\boldmath{\Lambda_c^+\rightarrow \Xi^- \pi^+ K^+}$ EVENT SELECTION}

\begin{figure}[ht]
  \centering\small
  \includegraphics[width=.44\textwidth]{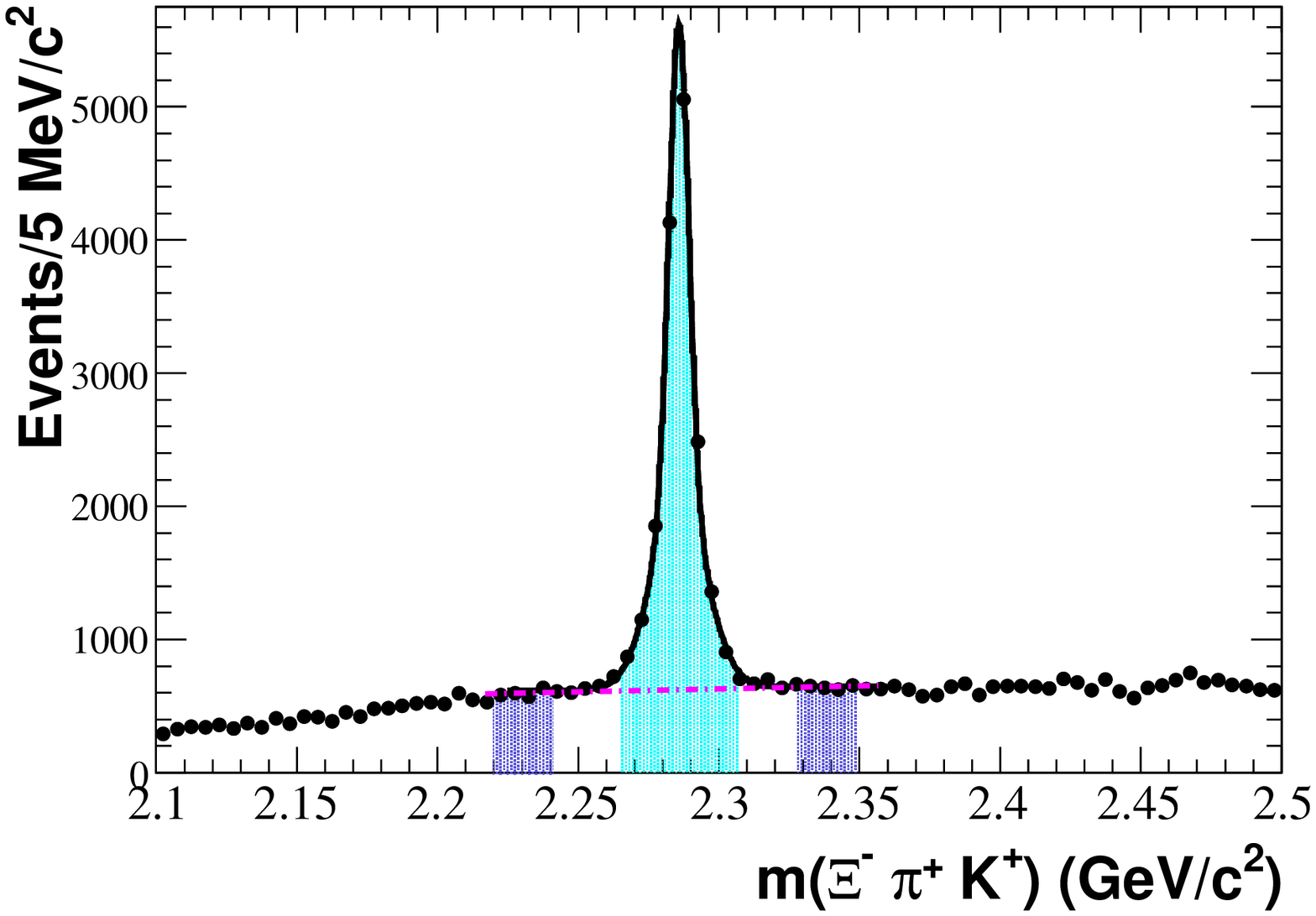}
  \includegraphics[width=.45\textwidth]{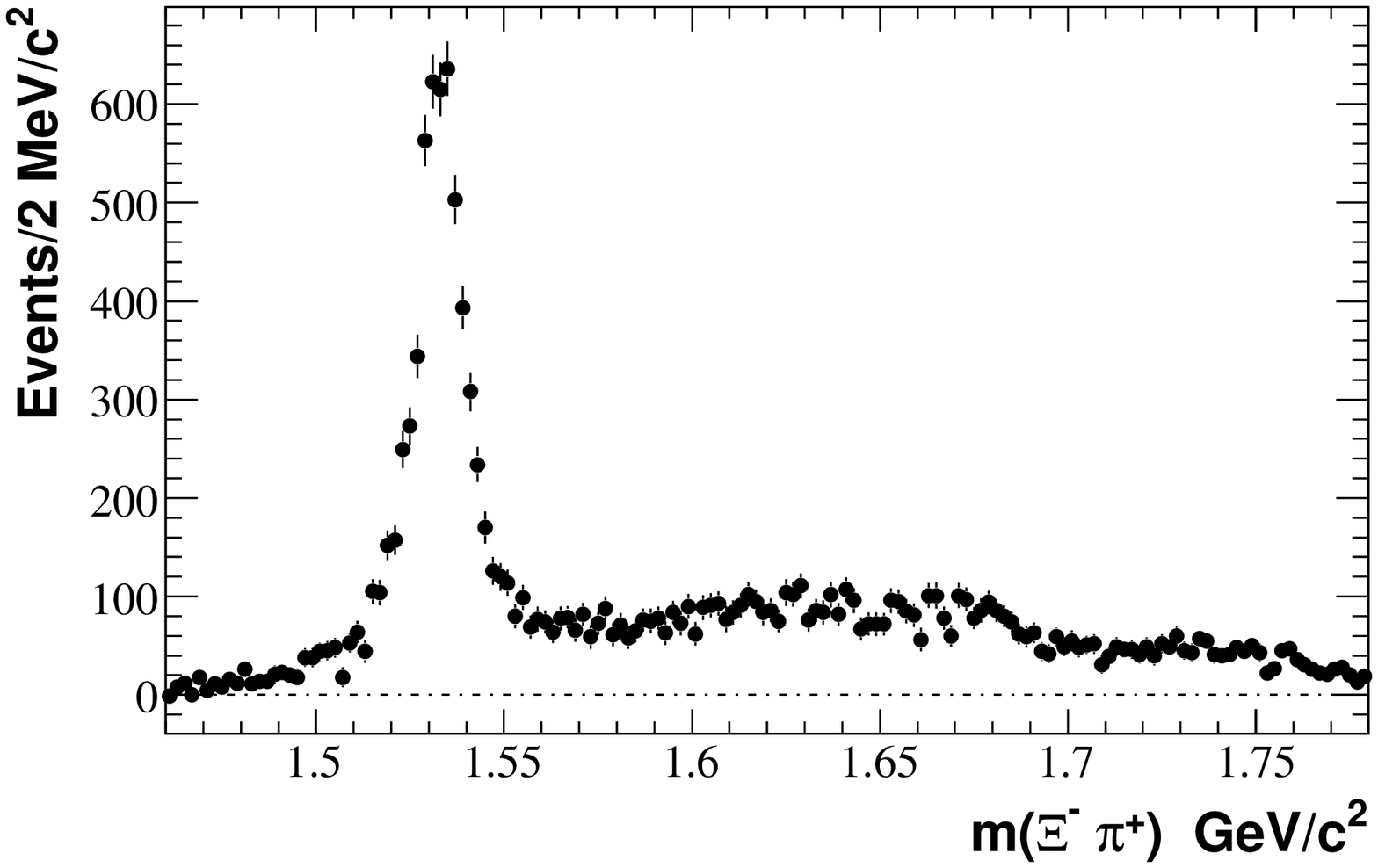}
  \begin{picture}(0.,0.)
    \put(-45,280){\bf{(a)}}
    \put(-46,115){\bf{(b)}}
    \end{picture}
  \caption{(a) The uncorrected $\Xi^- \pi^{+} K^{+}$ 
invariant mass  spectrum.  
The curve results from the fit described in the text.  The dot-dashed line indicates the fitted background contribution. 
The shaded areas delimit the signal (light area) and mass-sideband (dark area) regions.  
(b) The uncorrected $\Lambda_c^+$-mass-sideband-subtracted $\Xi^- \pi^{+}$ invariant mass projection for $\Xi^- \pi^{+} K^{+}$ candidates. 
In each figure, the points
with error bars represent the data.}
  \label{fig:mass}
\end{figure} 

The data sample used for this analysis was collected with the \babar\ detector
at the \pep2\ asymmetric-energy $e^+ e^-$ collider operating at center-of-mass (c.m.) energies 10.58 and
   10.54 GeV, and corresponds to a total integrated luminosity of
about 230 fb$^{-1}$.  

Charged particles are detected with a five-layer, double-sided
silicon vertex tracker (SVT) and a 40-layer drift chamber (DCH) with a
helium-isobutane gas mixture, placed in a 1.5-T solenoidal field produced
by a superconducting magnet. The  charged-particle momentum resolution
is approximately $(\delta p_T/p_T)^2 = (0.0013 \, p_T)^2 +
(0.0045)^2$, where $p_T$ is the transverse momentum   in GeV/c.
The SVT, with a typical
single-hit resolution of 10 $\mu$m,  measures the impact
parameters of charged-particle tracks in both the plane transverse to
the beam direction and along the collision axis.

Charged-particle types are identified from the ionization energy loss
(dE/dx) measured in the DCH and SVT, and from the Cherenkov radiation
detected in a ring-imaging Cherenkov device. Photons are
detected by a CsI(Tl) electromagnetic calorimeter  with an
energy resolution $\sigma(E)/E = 0.023\cdot(E/GeV)^{-1/4}\oplus
0.019$.

The return yoke of the superconducting coil is instrumented with
resistive plate chambers for the identification and muons and the
detection of neutral hadrons.
The detector is described
in detail in Ref.~\cite{ref:babar}.

The selection of $\Lambda_{c}^{+}$ candidates requires
the intermediate reconstruction of events consistent with $\Xi^- \rightarrow \Lambda \: \pi^-$ and 
$\Lambda \rightarrow p \: \pi^-$.  
Particle identification (PID) selectors 
based on specific energy loss (${\rm d}E/{\rm d}x$) and Cherenkov angle measurements are used to identify 
the proton, pion, and kaon final state tracks~\cite{ref:babar}. 
Each intermediate state candidate
is required to have invariant mass within a $\pm 3\sigma$ window centered on the fitted peak position of the relevant distribution,
where $\sigma$ is the mass resolution obtained from the fit.
A fit is then performed to the complete decay topology with the $\Lambda$ and $\Xi^-$ candidates constrained 
to their known mass values~\cite{PDG2006}.  The fit probability is required to be greater than 0.001 in order 
to ensure simultaneous satisfaction of the topological and mass constraint requirements; this reduces combinatorial 
background significantly and retains good signal efficiency.    
Since each weakly-decaying intermediate state (i.e., hyperon) is long-lived, 
 an improvement of the signal-to-background ratio is
  achieved by requiring that the decay vertex of each hyperon be
  displaced from its point of origin in the direction of its momentum
  vector.  
The distance between the  $\Xi^- K^+ \pi^+$ vertex and the  $\Xi^-$
decay vertex in the plane perpendicular to the collision axis must
exceed 1.5~mm in the $\Xi^-$ direction, and the distance between the $\Xi^-$ and $\Lambda$ decay
vertices must exceed 1.5~mm in the direction of the $\Lambda$ momentum vector.  
Finally, the momentum of the $\Lambda_c^+$ candidate in the $e^+ e^-$ c.m.
frame $p^*$ is required to be greater than $2.0$ GeV$/c$, 
since it is found empirically that this significantly reduces combinatorial background.  
The invariant mass spectrum of $\Lambda_{c}^{+}$ candidates which satisfy
these selection criteria 
before efficiency correction is shown in Fig.~\ref{fig:mass}(a).
A signal yield of $13035\pm 163$ events is obtained
from a fit which makes use of a signal function consisting of two Gaussians with a common center 
and a linear background function to the mass region 2.225 - 2.360 GeV/c$^2$.  The 
fit yields half-width-half-maximum 5.1 MeV/c$^2$ and has chi-squared per degree of
freedom ($\chi^2$/NDF) $19.6/20$.

\begin{figure}[ht]
  \centering\small
  \includegraphics[width=.45\textwidth]{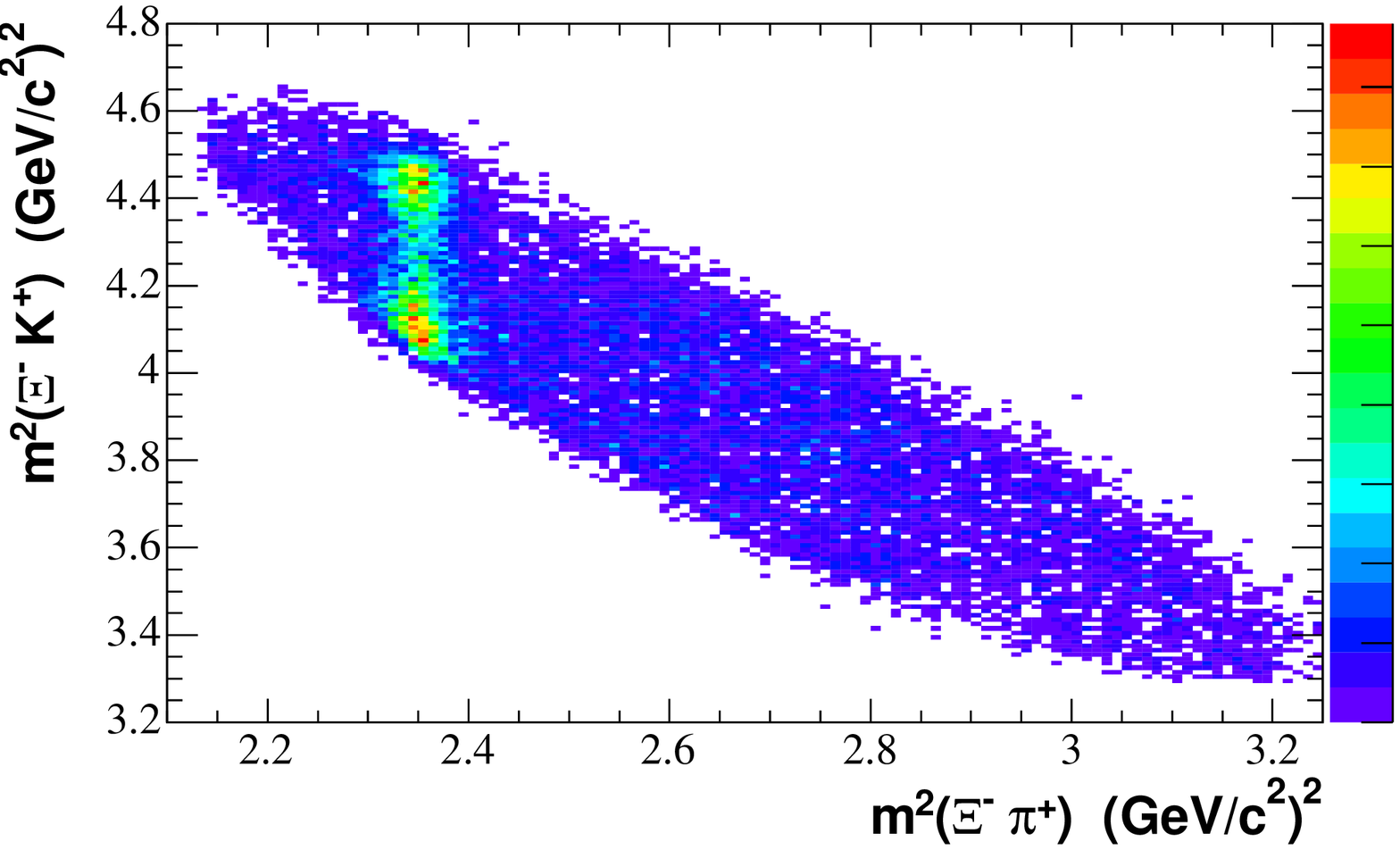}
  \includegraphics[width=.45\textwidth]{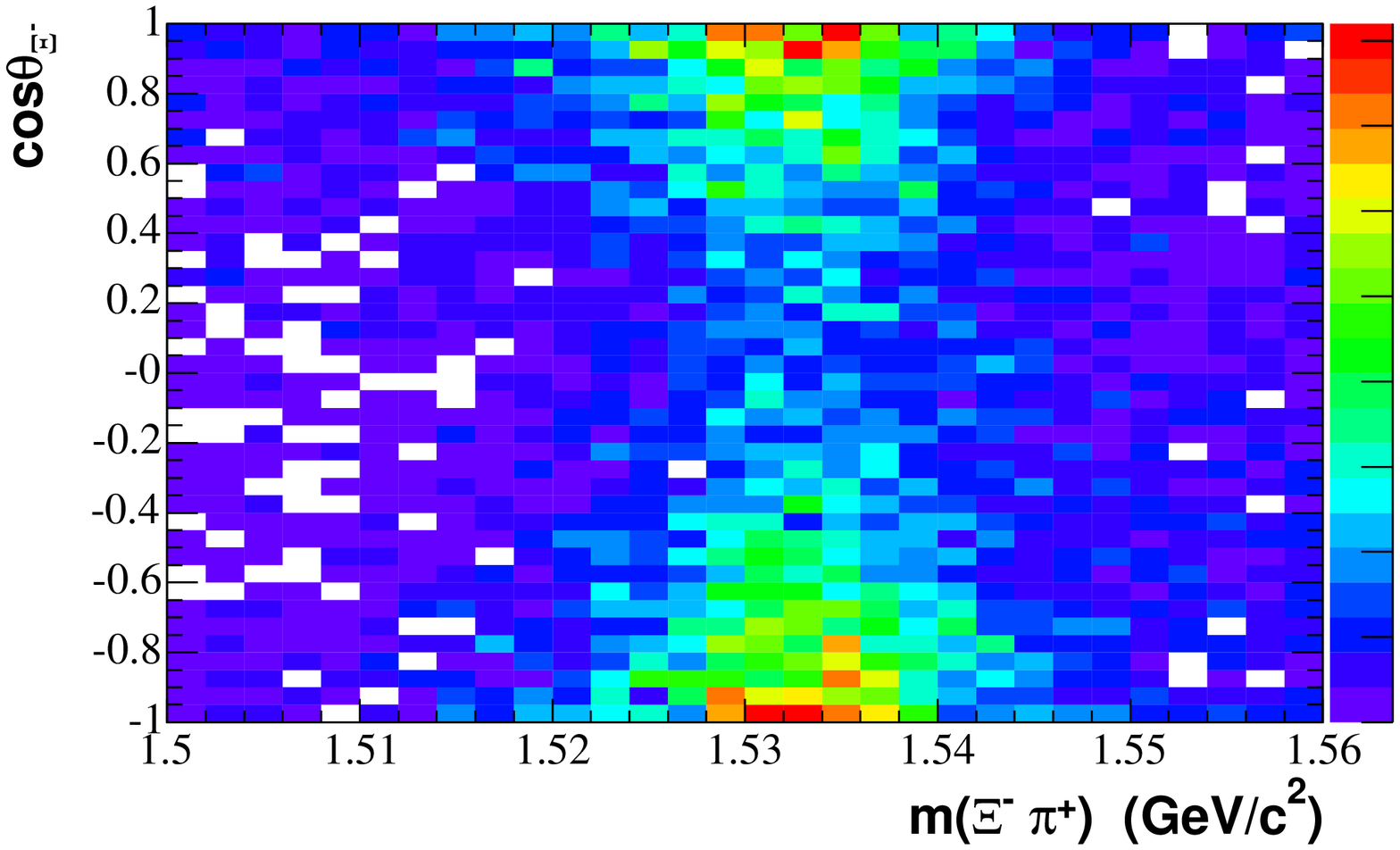}
  \begin{picture}(0.,0.)
    \put(-35,122){\bf\textcolor{white}{(b)}}
    \put(-35,280){\bf{(a)}}
    \end{picture}
  \caption{(a) The Dalitz plot of $\Xi^- K^{+}$ versus $\Xi^- \pi^{+}$ invariant mass-squared for the $\Lambda_c^+$ signal region.  
(b) The corresponding rectangular Dalitz plot for the $\Xi(1530)^0$ mass region. (The online version of this figure is in color.)}
  \label{fig:Dalitz}
\end{figure}

\section{$\boldmath{\Xi(1530)}$ SPIN MEASUREMENT}

\begin{figure}[hb]
  \centering\small
  \includegraphics[width=.385\textwidth]{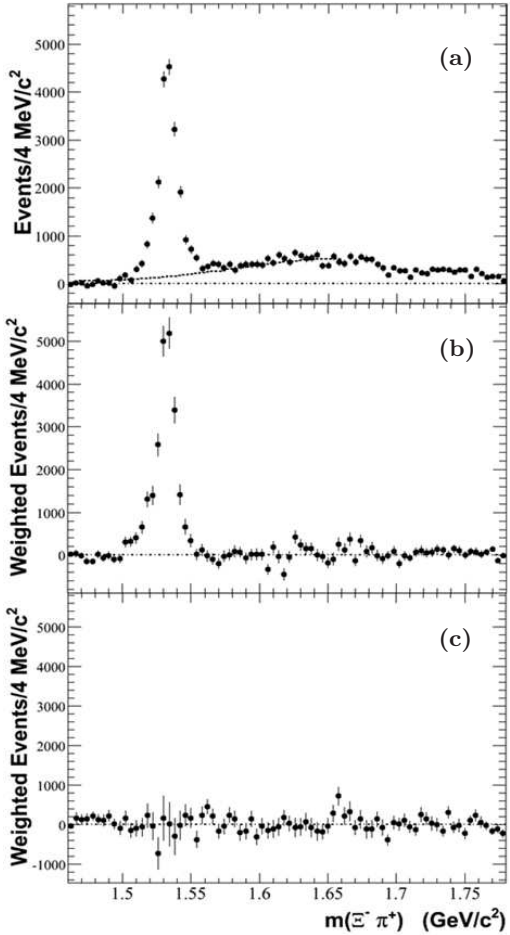}
  \begin{picture}(0.,0.)
   \put(-35,330){\bf{(a)}}
   \put(-35,220){\bf{(b)}}
   \put(-35,110){\bf{(c)}}
    \end{picture}
  \caption{The efficiency-corrected $\Lambda_c^+$-mass-sideband-subtracted 
(a) $\sqrt{2}P_0(\rm cos\,\theta_{\Xi^-})$,
 (b) $\sqrt{10}P_2(\rm cos\,\theta_{\Xi^-})$ and (c) $7/\sqrt{2}P_4(\rm cos\,\theta_{\Xi^-})$ 
moments of the $\Xi^{-} \pi^{+}$ system invariant mass distribution for the $\Lambda_c^+$ signal region.  
In (a) the dashed curve represents the estimated background contribution in the $\Xi(1530)$ region, obtained as 
described in the text.}
 \label{fig:XiPimoments}
\end{figure}

The Dalitz plot for $\Lambda_c^+\rightarrow \Xi^- \pi^+ K^+$ (Fig.~\ref{fig:Dalitz}) is dominated by the contribution from
$\Lambda_c^+\rightarrow \Xi(1530)^0 K^+$, where $\Xi(1530)^0\rightarrow \Xi^- \pi^+$ is a strong decay.
There is evidence for only one resonant structure, seen as the clear band at the nominal mass squared of
the $\Xi(1530)^0$.  
The background events in the signal region of Fig.1(a) are
represented by the events from the combined
sideband regions indicated in this figure, which correspond
to the same mass range~\cite{sig}. A corrected distribution associated
with the $\Lambda_c^+$ signal is obtained by subtraction (bin-by-bin) of the
corresponding distribution for the sidebands from that for
the signal region. This procedure is described as
``sideband subtraction'', and assumes linear mass dependence
of the background.
The sideband-subtracted projection of the $\Xi^- \pi^{+}$ invariant mass
for the $\Lambda_c^+$ signal region of Fig.~\ref{fig:mass} (a) is shown in Fig.~\ref{fig:mass} (b).

The helicity
formalism~\cite{ref:form, ref:form2} is applied to the quasi-two-body decay 
$\Lambda_c^+\rightarrow K^+ \Xi(1530)^0$ 
in order to examine the implications of various $\Xi(1530)^0$ spin hypotheses
for the angular distribution of the $\Xi^-$ from $\Xi(1530)^0$ decay, under the assumption 
that the $\Xi(1530)^0$ mass region is dominated by a single spin state.
As in Ref.~\cite{ref:omesp}, it is assumed that the spin of the charm baryon is 1/2.  
The choice of spin quantization axis along
the direction of the $\Xi(1530)^0$ in the charm baryon rest-frame (r.f.) has the result that 
the $\Xi(1530)^0$ inherits the spin
projection of the charm baryon, since any orbital angular momentum in the charm baryon decay has no projection in this direction.
It follows that, regardless of the spin $J$ of the $\Xi(1530)^0$, the density matrix which describes the $\Xi(1530)^0$ sample is
diagonal, with non-zero values only for the $\pm 1/2$ spin projection elements, 
i.e., the helicity $\lambda_i$ of the $\Xi(1530)^0$ can take only the values  $\pm 1/2$.  Since the final state $\Xi^-$ and $\pi^+$
have spin values 1/2 and 0, respectively, the net final state helicity $\lambda_f$ also can take only the values $\pm 1/2$.
The helicity angle $\theta_{\Xi^-}$ is defined as the angle between the direction of the $\Xi^-$ in the r.f. of the $\Xi(1530)^0$
and the quantization axis.  
Following the formalism of Ref.~\cite{ref:omesp}, the angular distribution of the $\Xi^-$ is then given by the total intensity,
\begin{eqnarray}
I\propto \sum_{\lambda_{i}, \lambda_{f}} \rho_{i\, i}\left |A^J_{\lambda_f}D^{J\, *}_{\lambda_{i} \lambda_{f}}(\phi, \theta_{\Xi^-}, 0)\right |^2,
\label{eq:he}
\end{eqnarray}
\noindent where the $\rho_{i\, i}$ ($i= \pm 1/2$) are the diagonal density matrix elements
inherited from the charm baryon, and the sum is over all initial and final helicity states.
The transition matrix element $A^J_{\lambda_f}$
represents the coupling of the $\Xi(1530)^0$ to the final state,
$D^{J}_{\lambda_{i} \lambda_{f}}$ is an element of the Wigner rotation matrix~\cite{ref:form3}, and the
$*$ denotes complex conjugation.   
The resulting $\Xi^-$ angular distribution integrated
over $\phi$ is obtained for spin hypotheses $J_{\Xi(1530)}=1/2$, $3/2,$ and $5/2$, respectively, as follows:
\begin{eqnarray}
{dN}/{d\rm{cos}\, \theta_{\Xi^-}}\hspace{-2mm}&\propto& 1+\beta\, \rm {cos}\,\theta_{\Xi^-}\\
\nonumber {dN}/{d\rm{cos}\, \theta_{\Xi^-}}\hspace{-2mm}&\propto& 1 + 3\,{\rm cos}^2\theta_{\Xi^-}+ \\ 
&& \hspace{-4mm} \beta\, \rm {cos}\theta_{\Xi^-}(5- 9\,\rm cos^2\theta_{\Xi^-}) \\
\nonumber {dN}/{d\rm{cos}\, \theta_{\Xi^-}}\hspace{-2mm}&\propto& 1-2\,{\rm cos}^2\theta_{\Xi^-}+5\,{\rm cos}^4\theta_{\Xi^-}+ \\
&&\hspace{-7mm}\beta\, \rm {cos}\,\theta_{\Xi^-}(5-26\,{\rm cos}^2\theta_{\Xi^-}+25\,{\rm cos}^4\theta_{\Xi^-}).
\end{eqnarray}
The coefficient of the asymmetric term,
$$ \beta=\left[\frac{\rho_{1/2\, 1/2}-\rho_{-1/2\, -1/2}}{\rho_{1/2\, 1/2}+\rho_{-1/2\, -1/2}}\right] \left[ {\frac{\left |A^J_{1/2}\right |^2-\left |A^J_{-1/2}\right |^2}{\left |A^J_{1/2}\right |^2+\left |A^J_{-1/2}\right |^2 }}\right ] , $$
is zero as a consequence of parity conservation in the strong decay of $\Xi(1530)^0$ to $\Xi^- \pi^+$, 
which implies $\left |A^J_{1/2}\right |=\left |A^J_{-1/2}\right |$. 
It should be noted that Eqs. 2-4 do not depend on any assumption as to
 the parity of the $\Xi(1530)$.  

The normalized angular distribution of the $\Xi^-$ obtained from Eq.~\ref{eq:he}, and 
expressed explicitly in Eqs. 2-4 for $J=1/2,3/2$ and 5/2, respectively, can be written in general as 
\begin{eqnarray}
\frac{dN}{d\rm{cos}\, \theta_{\Xi^-}}=N\left [ \sum_{{\it l}=0}^{{\it l}_{max}} \langle P_{\it l}\rangle P_{\it l}\left (\rm{cos}\, \theta_{\Xi^-}\right )\right ],
\end{eqnarray}
where ${\it l}_{max}=2J-1$, the value of each expansion coefficient $\langle P_{\it l}\rangle$ depends on $J$, 
and, if ${\it l}$ is odd, $\langle P_{\it l}\rangle=0.$
The Legendre Polynomials satisfy
\begin{eqnarray}
\int_{-1}^{1}d\rm{cos}\, \theta_{\Xi^-} {\it P_i}\left (\rm{cos}\, \theta_{\Xi^-}\right ) {\it P_j}\left (\rm{cos}\, \theta_{\Xi^-}\right )=\delta_{\it i\, j},
\end{eqnarray}
(i.e., ${\it P_l}\left (\rm{cos}\, \theta\right )=\sqrt{2\pi}Y^0_l\left (\rm{cos}\, \theta, \phi\right ),$ where $Y^0_l$ is a 
spherical harmonic function),
so that
\begin{eqnarray}
\int_{-1}^{1} \frac{dN}{d\rm{cos}\, \theta_{\Xi^-}}P_{\it l}\left (\rm{cos}\, \theta_{\Xi^-}\right ) d\rm{cos}\, \theta_{\Xi^-} 
&=&N\langle P_{\it l}\rangle. \end{eqnarray}
For a data distribution containing $N$ events, the left-hand side
 of Eq. 7 is approximately equal to 
\begin{eqnarray}
\nonumber \sum_{k=1}^{N}P_{\it l}\left (\rm{cos}\, \theta^{\it k}_{\Xi^-}\right ), 
\end{eqnarray}
since for large $N$, the sum over the observed events provides a good approximation to the integral; 
throughout this paper, this summation is termed ``the $P_{\it l}(\rm{cos}\,\theta_{\Xi^-})$ moment'' 
or simply `the $P_{\it l}$ moment'' of the data.  
Each assumption for $J$ defines ${\it l}_{max}$, so that $\langle P_{\it l}\rangle=0$ for ${\it l}>{\it l}_{max}$ and the 
$\langle P_{\it l}\rangle$ are calculable.  
For $J=1/2, 3/2, 5/2, $ $l_{max}=0, 2, 4$, with the corresponding $\langle P_{\it l_{max}}\rangle$ 
values $1/\sqrt{2}$, $1/\sqrt{10}$ and $\sqrt{2}/7$, respectively. 
The relation
\begin{eqnarray}
\sum_{k=1}^{N}\frac{P_{\it l_{max}}\left (\rm{cos}\, \theta^{\it k}_{\Xi^-}\right )}{\langle P_{\it l_{max}}\rangle }\approx N
\end{eqnarray}
implies that the number
of $\Xi(1530)^0$ signal events in a given mass interval
 is well-approximated if each event $k$ is given weight
\begin{eqnarray}
w_k = \frac{P_{\it l_{max}}\left (\rm{cos}\, \theta^{\it k}_{\Xi^-}\right )}{\langle P_{\it l_{max}}\rangle },
\end{eqnarray}
after efficiency correction~\cite{eff} and background subtraction. 
                                                             
Since the angular distribution shown in Fig.~\ref{fig:Dalitz}(b) is clearly not flat, $\Xi(1530)$ spin 1/2 is ruled out.  
In order to test the $J=3/2$ $(5/2)$ hypothesis, each event is given a weight $w_k = \sqrt{10} P_2(\rm{cos}\,\theta^{\it k}_{\Xi^-})$ 
($\frac{7}{\sqrt{2}} P_4(\rm{cos}\,\theta^{\it k}_{\Xi^-})$).
Figure~\ref{fig:XiPimoments}(a) shows the distribution of the $\sqrt{2}P_0(\rm cos\,\theta_{\Xi^-})$ moment 
which is just the efficiency-corrected distribution corresponding to Fig.~\ref{fig:mass}(b) (the average efficiency is $\sim 27$\%).
The $\sqrt{10}P_2(\rm cos\,\theta_{\Xi^-})$ and  $7/\sqrt{2}P_4(\rm cos\,\theta_{\Xi^-})$ moments are shown in 
Figs.~\ref{fig:XiPimoments}(b) and~\ref{fig:XiPimoments}(c), respectively. 
Figure~\ref{fig:XiPimoments}(b) indicates that spin 3/2 is strongly favored, as essentially all of the $\Xi(1530)$ signal is retained.
In contrast, the $7/\sqrt{2}P_4(\rm cos\,\theta_{\Xi^-})$ moment shown in Fig.~\ref{fig:XiPimoments}(c) 
is consistent with zero in the $\Xi(1530)$ signal region, so that spin 5/2 is clearly ruled out.  
The results for $l_{max}\ge 6$ (not shown) are similar to those of Fig~\ref{fig:XiPimoments}(c).  
In order to quantify these results, the region $1.50\leq m(\Xi^- \pi^+)\leq 1.65$ GeV/c$^2$ is defined as the $\Xi(1530)$ signal 
region.  The dashed curve of Fig.~\ref{fig:XiPimoments}(a) corresponds to a fit 
to the region $m(\Xi^- \pi^+)\leq 1.66$ GeV/c$^2$ with the signal region excluded;  
the fit function is a third order polynomial multiplied by phase space.   
This yields an estimated signal of $19159\pm 581$ events.  
For Figs.~\ref{fig:XiPimoments}(b) and~(c), the moment sums for the signal region are $23355\pm 894$ and 
$78\pm 1410$, respectively.  Clearly, $J=3/2$ is the only viable $\Xi(1530)$ spin value.  
It follows that, based on the results of Refs.~\cite{ref:schlein, ref:button} (i.e., $J^P=3/2^+$ or $J^P=5/2^-$),  
the present analysis, which shows that $J=3/2$, also establishes positive parity, and that the $\Xi^-\pi^+$ system which results 
from the decay is in a $P$-wave orbital angular momentum state.    
Here, and in Refs.~\cite{ref:schlein,ref:button}, it is assumed that the $\Xi^-$ 
has positive parity~\cite{PDG2006}.  

\begin{figure}[hb]
  \centering\small
\includegraphics[width=.45\textwidth]{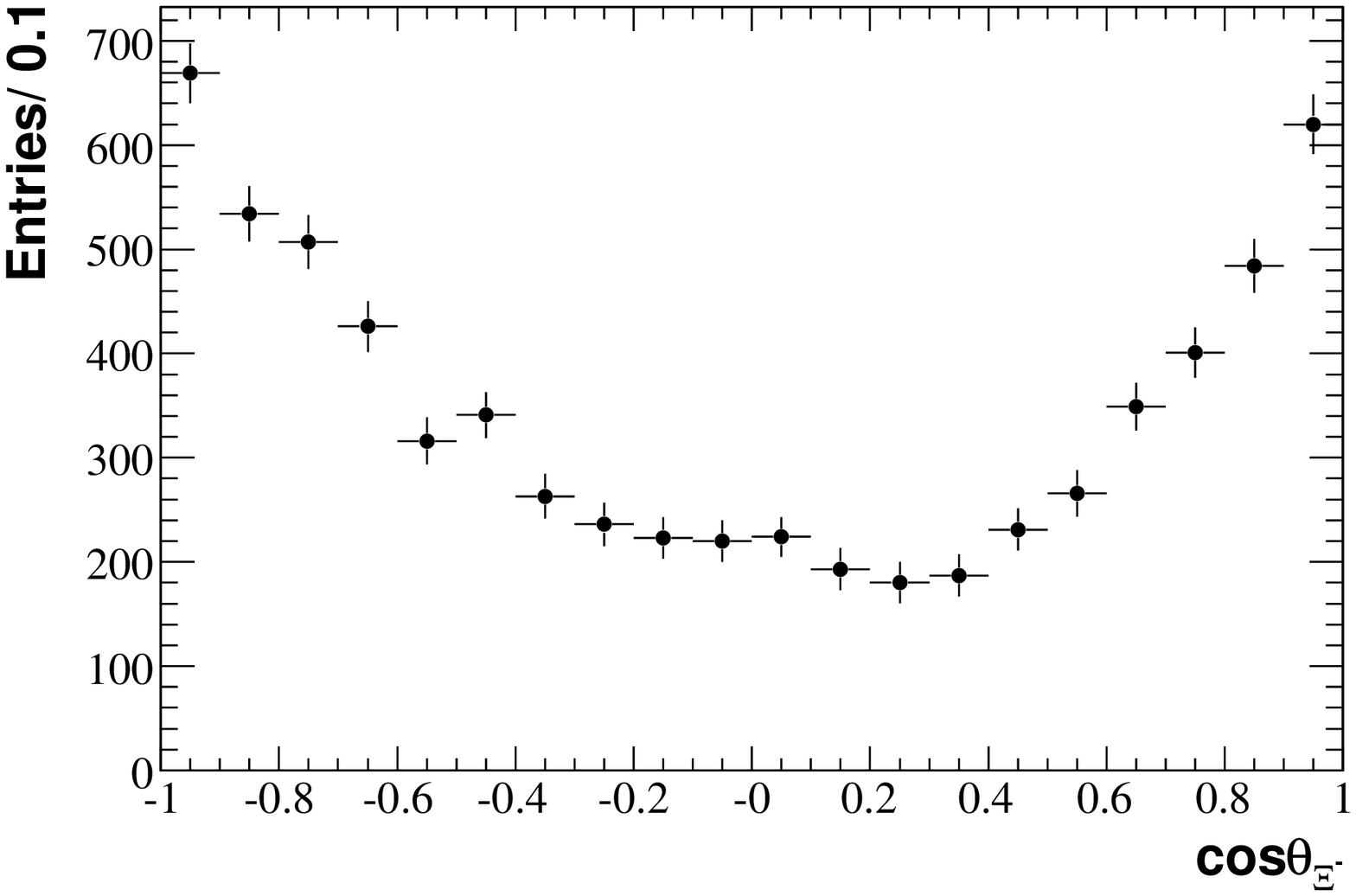}
  \includegraphics[width=.45\textwidth]{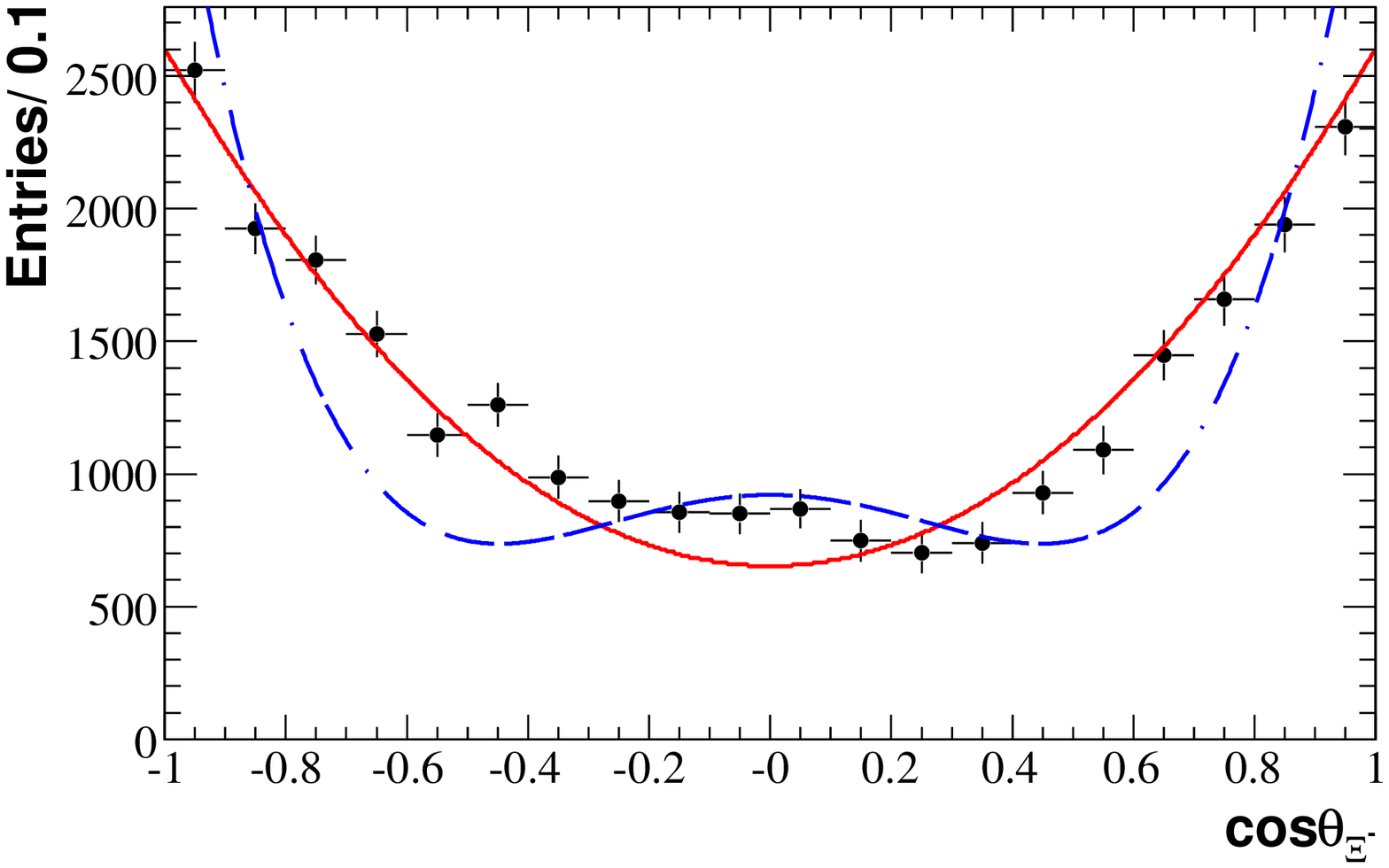}
  \begin{picture}(0.,0.)
    \put(-65,282){\bf{(a)}}
    \put(-65,119){\bf{(b)}}
    \end{picture}
  \caption{The cos$\,\theta_{\Xi^-}$ distribution for $\Lambda_c^+\rightarrow \Xi^- \pi^+ K^+$ data 
 in the $\Xi(1530)^0\rightarrow \Xi^- \pi^+$ 
signal region (a) before, and (b) after, efficiency correction.  
The solid (dashed) curve corresponds to the parametrization of the $\Xi(1530)$ angular distribution for
the assumption of {\it pure} spin 3/2 (5/2).}
  \label{fig:HelicityDis}
\end{figure}

\section{THE $\boldmath{\Xi(1530)}$ MASS REGION}

Although Fig.~\ref{fig:XiPimoments} clearly establishes spin 3/2 for the $\Xi(1530)$, the analysis of the $\Xi^- \pi^+$ system 
described in the remainder of this paper indicates that a detailed understanding of the data is much less straightforward than 
Fig.~\ref{fig:XiPimoments} might indicate.  
If the momentum of the $\Xi^-$ in the $\Xi^-\pi^+$ r.f. is denoted by $q$, a 
Breit-Wigner (BW) amplitude corresponding to orbital angular momentum $L$ should be proportional to the centrifugal barrier 
factor $q^L$~\cite{blatt}.  The lineshape for a $\Xi^-\pi^+$ $P$-wave BW should then be skewed toward high mass values.  
However, the distribution of Fig.~\ref{fig:XiPimoments}(b), which should represent the square of a $P$-wave amplitude, appears 
to be skewed toward low mass values, in contradiction of this expectation. 
Furthermore, if the distribution in Fig.~\ref{fig:XiPimoments}(a) is considered to 
represent a sum of squares of $\Xi^- \pi^+$ amplitudes, for which that in Fig.~\ref{fig:XiPimoments}(b) represents the $J=3/2$ contribution, their 
difference would be expected to behave like the background distribution in Fig.~\ref{fig:XiPimoments}(a) in the $\Xi(1530)^0$ region.  However, the 
$\Xi(1530)$ signal in Fig.~\ref{fig:XiPimoments}(b) contains $\sim 4000$ events more than that in  
Fig.~\ref{fig:XiPimoments}(a), as indicated above, so that when the former is subtracted from the latter, the residual 
distribution exhibits a strong dip in the $\Xi(1530)$ region, and even reaches negative values.  This behavior is clearly at odds 
with a simple interpretation of these distributions.  

Moreover, the cos$\,\theta_{\Xi^-}$ distribution in the $\Xi(1530)^0$ signal
region indicates that a description in terms of a single $\Xi^- \pi^+$
amplitude corresponding to a resonant structure is an over-simplification.  
The $\Lambda_c^+$ mass-sideband-subtracted cos$\,\theta_{\Xi^-}$ distribution for the $\Xi(1530)^0$ signal
region (Fig.~\ref{fig:HelicityDis}) exhibits a 
predominantly quadratic behavior, which indicates clearly that the spin of the
 $\Xi(1530)$ is not 1/2. A function $\propto \left( 1 + 3\rm cos^2\theta\right )$ (solid curve of Fig.~\ref{fig:HelicityDis}(b), the 
parametrization of Eq.~3) which corresponds to $J=3/2$ for the $\Xi(1530)^0$ fits the data best,
although there are clear deviations from the curve, and the fit confidence level (c.l.) is only 0.0003.
The fit with the parametrization which corresponds to $J=5/2$ (dashed curve, Eq.~4) is extremely poor, with c.l. $6\times 10^{-44}$,
as would be expected from the projection of Fig.~\ref{fig:XiPimoments}(c).
In addition, the distribution of Fig.~\ref{fig:HelicityDis}(b) exhibits signs of forward-backward asymmetry.
                                                                                                                                                 
The above symptoms indicate that a more complicated description is required if a quantitative
understanding of the $\Xi^- \pi^+$ system is to be achieved.

\begin{figure}[!t]
  \centering\small
  \includegraphics[width=.5\textwidth]{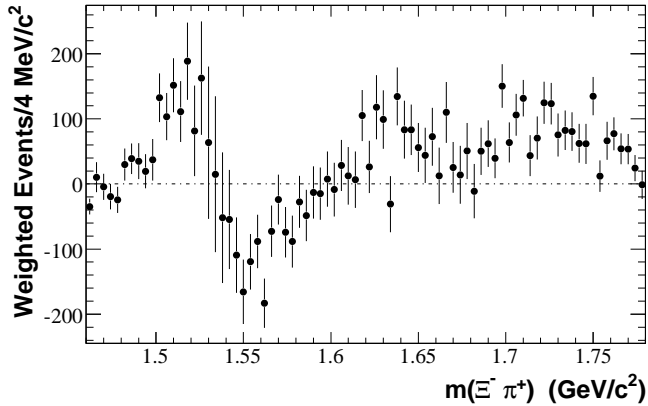}
  \begin{picture}(0.,0.)
    \end{picture}
  \caption{The efficiency-corrected $P_1(\rm cos\,\theta_{\Xi^-})$ moment of the $\Xi^{-} \pi^{+}$ system invariant mass distribution
corresponding to the $\Lambda_c^+$ signal region.  
The distributions for the sideband regions are consistent with zero, and so are not subtracted.
}
 \label{fig:P1moment}
\end{figure}

Strong interactions in the $\Xi^- \pi^+$ system may give rise to interference
between the resonant $P$-wave  $\Xi(1530)$ amplitude and other
$\Xi^- \pi^+$ amplitudes.  Evidence for interference is seen in the behavior of
the $P_1(\rm cos\,\theta_{\Xi^-})$ moment of the $\Xi^{-} \pi^{+}$ system as a function of invariant mass.
The distribution shown in Fig.~\ref{fig:P1moment} is consistent with the interference pattern resulting from the
rapid oscillation due to $\Xi(1530)$ $P$-wave BW phase motion
in the presence of an amplitude with small slowly varying relative phase; 
the projection would then approximate the real part of the BW amplitude, as observed.  
 
The oscillatory pattern seen in Fig.~\ref{fig:P1moment} 
is not observed for the high and low $\Lambda_c^+$ mass sideband regions, 
which confirms that the pattern is indeed due to $\Xi(1530)$ phase-motion in events produced from
signal $\Lambda_c^+$ candidates, and is not simply an artifact of combinatorial background.  
As mentioned above, the $P_1(\rm cos\,\theta_{\Xi^-})$ moment for $m(\Xi^- \pi^+)<1.58$ GeV/c$^2$ behaves very much like the real part of
the $\Xi(1530)$ BW amplitude, which suggests that the phase of the amplitude yielding the
interference effect is close to zero.  The proximity of $\Xi^-\pi^+$ threshold, and the fact that the interference
is seen in the $P_1(\rm cos\,\theta_{\Xi^-})$ moment, suggest that the effect is due primarily to
the presence of an $S$-wave $\Xi^- \pi^+$ amplitude.

If it is assumed that only total spin $J$=1/2 and 3/2 amplitudes contribute,   
and if the description is restricted to $S$, $P$ and $D$ waves, 
the following angular distribution for the
$\Xi^-$ produced in $\Xi^{-} \pi^{+}$ decay is obtained:
\begin{eqnarray}
\frac{dN}{d\rm{cos}\,\theta_{\Xi^-}}&=&\nonumber \left[\frac{1}{2}|S^{1/2}|^2+|P^{3/2}|^2\left(\frac{1+3\rm{cos}^2\theta_{\Xi^-}}{4}\right)\right .\\
\nonumber&+&\left .\sqrt{2}Re\left ( S^{1/2}P^{3/2 *}\right )\rm{cos}\,\theta_{\Xi^-}\right] \\
\nonumber&+& \left[\frac{1}{2}|P^{1/2}|^2+|D^{3/2}|^2\left(\frac{1+3\rm{cos}^2\theta_{\Xi^-}}{4}\right)\right . \\
&+&\left . \sqrt{2}Re\left ( P^{1/2}D^{3/2 *}\right )\rm{cos}\,\theta_{\Xi^-}\right]
\label{eq:sp}
\end{eqnarray}
where the amplitude notation is $L^J$.

The angular structure associated with the $J=L+1/2$ terms (Eq.~10, first bracket) is identical to that associated 
with the $J=L-1/2$ terms (Eq.~10, second bracket), i.e., there is a Minami ambiguity~\cite{minami}.
It follows that there are more unknown quantities than measurables, so that a complete set
 of amplitudes cannot be extracted from the data.  
However, since the $\Xi(1530)$ is a $P^{3/2}$ resonance, it is reasonable to attribute the $P_1(\rm cos\,\theta_{\Xi^-})$ 
moment behavior of Fig.~\ref{fig:P1moment} to the $S^{1/2}$-$P^{3/2}$ interference term of Eq.~10; in addition, 
$D$-wave amplitudes would not be expected to be significant for $\Xi^- \pi^+$ mass values close to threshold, so that 
a simple model which incorporates only $S^{1/2}$ and $P^{3/2}$ amplitudes might describe the data.  This would imply that the 
intensity distribution of Fig.~\ref{fig:XiPimoments}(b) corresponds to $|P^{3/2}|^2$ only.  However,  
as discussed above, the difference in the distributions 
of Figs.~\ref{fig:XiPimoments}(a) and~(b) dips strongly in the $\Xi(1530)$ region, even reaching 
negative values, and so cannot be described by $|S^{1/2}|^2$. 
This indicates that the data in the $\Xi(1530)$ mass region
require a more complicated explanation.

\section{THE $\boldmath{\Xi^- \pi^+}$ SYSTEM AT HIGHER MASS}

The inclusion of a $D^{3/2}$ contribution (Eq.~10) does not solve the problem of the $\Xi(1530)$ mass region described
at the end of Section IV, since Fig.~\ref{fig:XiPimoments}(b) 
then corresponds to $|P^{3/2}|^2 + |D^{3/2}|^2$ and Fig.~\ref{fig:XiPimoments}(a) to 
$|S^{1/2}|^2 + |P^{3/2}|^2 + |D^{3/2}|^2$.  
If the model is extended to include a $D^{5/2}$ amplitude, 
the Legendre Polynomial moments, $P_0$--$P_4$, are
 expressed in terms of the amplitudes as follows:
\begin{eqnarray}
\nonumber  P_0 &=&\frac{1}{\sqrt{2}}\left [\left |S^{1/2}\right |^2+\left |P^{1/2}\right |^2 + \left |P^{3/2}\right |^2\right . \\
                  && \left . +\left |D^{3/2}\right |^2 +\left |D^{5/2}\right |^2\right ]\\
\nonumber  P_1 &=& \frac{2}{\sqrt{3}}\left [ Re\left( S^{1/2} P^{3/2\,*}\right ) + Re\left( P^{1/2} D^{3/2\,*}\right )\right ] \\
                 && +\frac{6}{5}Re\left( P^{3/2} D^{5/2\,*}\right )  \\
\nonumber  P_2 &=&\frac{1}{\sqrt{10}}\left [ \left |P^{3/2}\right |^2+\left |D^{3/2}\right |^2+\frac{8}{7}\left |D^{5/2}\right |^2\right . \\
                 && \left . +\sqrt{20}Re\left( S^{1/2} D^{5/2\,*}\right ) \right ] \\
 P_3 &=& \frac{4}{5}\sqrt{\frac{3}{7}}Re\left( P^{3/2} D^{5/2\,*}\right ) \\
 P_4 &=& \frac{\sqrt{2}}{7}\left |D^{5/2}\right |^2
\end{eqnarray}

 These five equations involve nine unknown quantities (five amplitude
 magnitudes and four relative phase angles), and so cannot be solved.
 Additional input from polarization moments is required. Such an
 analysis is beyond the scope of the present paper.
 If we assume that the $P^{1/2}$ and $D^{3/2}$ amplitudes
 can be ignored, Eqs. 11-15 can be solved in principle. However, as is
 discussed below, even such a simplified model encounters difficulties in
 the $\Xi(1530)$ region.

\begin{figure}[ht]
  \centering\small
  \includegraphics[width=.5\textwidth]{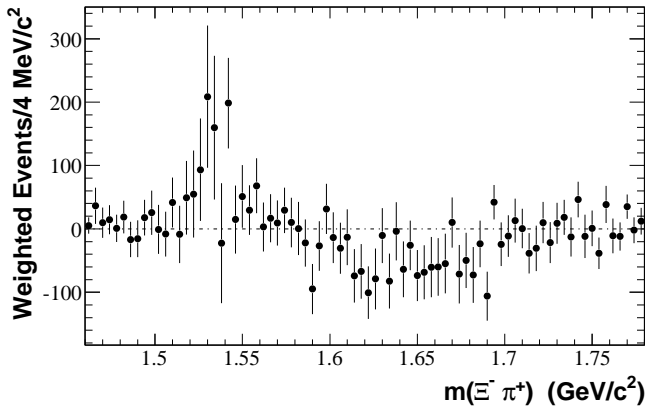}
\begin{picture}(0.,0.)
    \end{picture}
  \caption{The efficiency-corrected $P_3(\rm cos\,\theta_{\Xi^-})$
moment of the $\Xi^{-} \pi^{+}$ invariant mass distribution
for the $\Lambda_c^+$ signal region.
The distributions for the sideband regions are consistent with zero, and so are not subtracted.
}
 \label{fig:P3moment}
\end{figure}

\begin{figure}[ht]
  \centering\small
  \includegraphics[width=.5\textwidth]{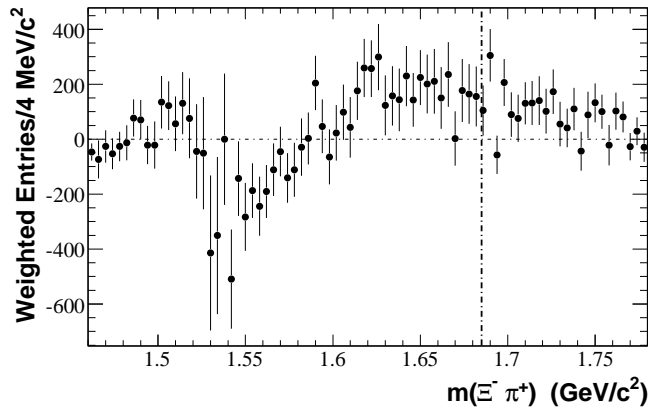}
  \begin{picture}(0.,0.)
    \end{picture}
  \caption{The efficiency-corrected $P_1 - (\sqrt{21}/2) P_3$
moment of the $\Xi^{-} \pi^{+}$ system invariant mass distribution, corresponding to the $\Lambda_c^+$ signal region.  
The dot-dashed line indicates the $\Xi(1690)^0$ mass value~\protect\cite{PDG2006}.  
The distributions for the sideband regions are consistent with zero, and so are not subtracted.
}
 \label{fig:P1P3Dif}
\end{figure}

  In the context of this model, the absence of any significant $P_4$ moment
 in Fig.~\ref{fig:XiPimoments}(c) indicates via Eq.~15 that $|D^{5/2}|$ must be small. However,
 since $P^{3/2}$ is large, $P^{3/2} - D^{5/2}$ interference might be seen in 
the mass-dependence of the $P_3$ moment (Eq.~14). This is shown in Fig.~\ref{fig:P3moment}, where small, but
 significant, deviations from zero are in fact observed. Since there is
 a $P^{3/2} - D^{5/2}$ interference contribution to Eq.~12, an improved
 measure of the mass dependence of $S^{1/2} - P^{3/2}$ interference is
 obtained by subtracting ($\sqrt{21}/2) P_3$ from $P_1$. The 
 $P_1 - (\sqrt{21}/2)  P_3$ distribution is shown in Fig.~\ref{fig:P1P3Dif}, 
 and the dip in the mass region $1.63 - 1.70$ GeV/c$^2$ of Fig.~\ref{fig:P1moment} has been removed by
 this procedure. Before the $\Xi(1530)$ region is examined in more
 detail, the behavior of $P_1 - (\sqrt{21}/2) P_3$ is considered in the mass
 region above $\sim 1.6$ GeV/c$^2$.

\begin{figure}[ht]
  \centering\small
  \includegraphics[width=.5\textwidth]{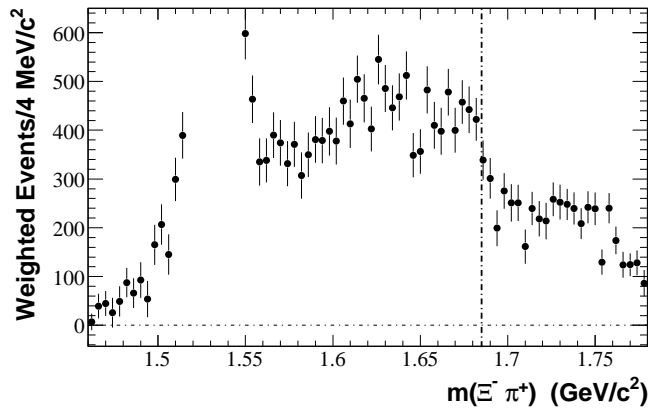}
  \begin{picture}(0.,0.)
    \end{picture}
  \caption{The efficiency-corrected $\Lambda_c^+$ mass-sideband-subtracted $P_0(\rm cos\,\theta_{\Xi^-})$
moment of the $\Xi^{-} \pi^{+}$ system invariant mass distribution for the $\Lambda_c^+$ signal region
(the distribution of Fig.~\protect\ref{fig:XiPimoments}(a) with the $\Xi(1530)$ region suppressed).
The vertical dot-dashed line indicates the $\Xi(1690)^0$ mass value~\protect\cite{PDG2006}.}
 \label{fig:P0Zoom}
\end{figure}

\begin{figure}[ht]
  \centering\small
  \includegraphics[width=0.35\textwidth]{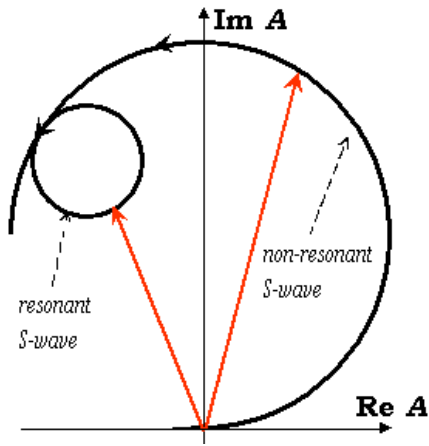}
  \caption{A cartoon of an Argand diagram which 
illustrates a possible cause for the dip in the $\Xi^- \pi^+$ invariant mass distribution
 due to the presence of the $\Xi(1690)^0\rightarrow \Xi^- \pi^+$.}
 \label{fig:Argand}
\end{figure}

  It is interesting to consider this in comparison to the distribution
 of Fig.~\ref{fig:XiPimoments}(a) with the $\Xi(1530)$ region suppressed (Fig.~\ref{fig:P0Zoom}), which
 shows a significant decrease in intensity at $\sim 1.7$ GeV/c$^2$. As mentioned
 previously, the behavior of the $P_1$ moment in the $\Xi(1530)$ region
 indicates a small $S^{1/2}$ amplitude with phase $\sim 0$ deg. relative to
 the $P^{3/2}$ amplitude in order that $P_1$ closely resemble the real
 part of the $\Xi(1530)$ BW amplitude. If the $S^{1/2}$ amplitude
 did not change significantly at higher mass values, the BW amplitude
 would cause $P_1$ to approach zero from below with increasing mass.
 Instead, $P_1$ passes through zero at $\sim 1.6$ GeV/c$^2$ and remains positive
 thereafter (Fig.~\ref{fig:P1P3Dif}). Since $P_1$ represents the projection of the
 $S^{1/2}$ amplitude onto the $P^{3/2}$ amplitude, this means that the
 $S^{1/2}$ phase is increasing significantly, is only 90 deg. behind the
 $P^{3/2}$ phase at $\sim 1.6$ GeV/c$^2$ where $P_1$ is $\sim 0$, and continues to
 increase at higher mass. The dip in the mass distribution of Fig.~\ref{fig:P0Zoom}
 is in the vicinity of the $\Xi(1690)$, which is known to have a
 small coupling to $\Xi^- \pi^+$~\cite{adamovich}.
This dip could occur as the result of the coherent addition of a small, resonant
 $\Xi(1690)$ amplitude to the slowly increasing $S^{1/2}$ amplitude,
 as shown schematically by the cartoon of Fig.~\ref{fig:Argand}. Here the large circle
 represents a slowly-varying non-resonant $S^{1/2}$ amplitude for which
 the phase reaches 90 deg. at mass $\sim 1.6$ GeV/c$^2$, relative to a $P^{3/2}$
 amplitude; the latter should be approximately aligned with the negative real axis
 at this mass value. As the phase increases beyond 90 deg., the $S^{1/2}$
 projection on the $P^{3/2}$ amplitude [i.e., $P_1$] will increase as
 seen in Fig.~\ref{fig:P1P3Dif}. The small circle represents the subsequent coherent
 addition of a small resonant $\Xi(1690)$ amplitude.  The
 resultant amplitude will then yield a dip in overall intensity in the
 $\Xi(1690)$ region with very little effect on the phase, and hence on
 $P_1$ (cf. Fig.~\ref{fig:P1P3Dif}). The inference which can be drawn is that the $\Xi(1690)$ decays strongly to
 the $\Xi^- \pi^+$ system in an $S$-wave orbital state, and hence that it
 has spin-parity $1/2^{-}$. As such, this represents the first
 experimental information on the spin-parity of the $\Xi(1690)$.
 Spin 1/2 is favored also by an analysis of the
 Dalitz plot corresponding to the decay process $\Lambda_c^+ \rightarrow \Lambda \bar{K}^0 K^+$~\cite{thesis}.
 
  The behavior of the $S^{1/2}$ amplitude described above is remarkably
 similar to that obtained for the $I=1/2$ $S$-wave $K^- \pi^+$ scattering
 amplitude in the LASS experiment~\cite{lass}. There the slow, monotonic
 increase in the $S$-wave amplitude at low mass is described by an
 effective range parametrization. The phase reaches $\sim 90$ deg. before
 the coherent addition of a resonant $K^*_0(1430)$ contribution takes
 effect, and the resultant amplitude decreases quickly almost to zero.
 The main difference in the $K^- \pi^+$ case is that the resultant amplitude
 remains elastic (within error) up to $K \eta'$ threshold, so that the
 decrease in $S$-wave intensity is quite substantial. Since the
 $\Xi(1690)$ decays significantly via modes other than $\Xi^- \pi^+$,
 a similar mechanism would be expected to yield less dramatic results, as is in fact observed in Fig.~\ref{fig:P0Zoom}.
This similarity between $K \pi$ and $\Xi \pi$ scattering amplitudes may be an example 
of the proposed effective supersymmetry between mesons and baryons involving 
the replacement of a light anti-quark in the meson by a light diquark to form 
the related baryon~\cite{Lipkin}.  For $I=1/2$, the amplitudes describing 
$K^- \pi^+$ scattering are the same as for $K^+ \pi^-$, and similarly those 
describing $\Xi^- \pi^+$ scattering are proportional to those for $\Xi^0 \pi^-$ scattering; 
$K^+\pi^-$ scattering is then converted to $\Xi^0\pi^-$ scattering by replacing the $\bar{s}$ quark 
in the $K^+$ with an $(ss)$ diquark to obtain the $\Xi^0$.  In Ref.~\cite{Lipkin}, the 
effective symmetry is demonstrated by relating various baryon-baryon and meson-meson mass 
differences with impressive precision.  It seems reasonable to conjecture 
that this symmetry might also be manifest in the dynamics of appropriately related meson-meson 
and baryon-meson scattering processes.

\section{THE $\boldmath{\Xi(1530)^0}$ LINESHAPE}

  In the $\Xi(1530)$ region,
 $S^{1/2} - D^{5/2}$ interference does in fact contribute to the $P_2$
 moment distribution in Fig.~\ref{fig:XiPimoments}(b) (cf.~Eq.~13), but not to the
 distribution in Fig.~\ref{fig:XiPimoments}(a) (cf.~Eq.~11), so that the $\Xi(1530)$ signal
 in Fig.~\ref{fig:XiPimoments}(b) might be larger than that in Fig.~\ref{fig:XiPimoments}(a). In addition, this
 contribution might distort the lineshape in Fig.~\ref{fig:XiPimoments}(b), but should not
 affect that in Fig.~\ref{fig:XiPimoments}(a), which is obtained by integration over
 cos$\,\theta_{\Xi^-}$. In order to test this conjecture, fits to the
 distributions in Figs.~\ref{fig:XiPimoments}(a) and~\ref{fig:XiPimoments}(b) are performed in which the
 $\Xi(1530)$ lineshape is described by
\begin{eqnarray}
\frac{dN}{dm} &=&C\frac{\left (p^{2L}/ D_L(p,R)\right)\left ( q^{2l}/D_l(q,R)\right ) }{\left (m_0^2 - m^2\right )^2 + m_0^2 \Gamma_{tot}(m)^2 }p\cdot q ,
\end{eqnarray}
 where $C$ is a constant, $p$ is the momentum of the $K^+$ in the $\Lambda_c^+$
 r.f., and $q$ is the momentum of the $\Xi^-$ in the
 $\Xi^- \pi^+$ r.f.; $L$ is the orbital angular momentum in the $\Lambda_c^+$
 decay ($L=1$ is chosen), and $l$ that in the $\Xi(1530)$ decay (for
 which $l=1$); $D_L$, $D_l$ are Blatt-Weisskopf barrier factors~\cite{blatt} with
 radius parameter $R$, and, for example, $D_1(q,R) = 1+(qR)^2$; $R=3$ GeV$^{-1}$ ($\sim 0.6$ f) is chosen~\cite{lass}.   
The $\Xi(1530)$ mass is $m_0$, and $\Gamma_{tot}(m)$
 its mass-dependent total width, which consists of the sum of partial
 widths to $\Xi^0 \pi^0$ and $\Xi^- \pi^+$. If the mass
 differences between these modes are ignored, the mass-dependent total width is then 
\begin{eqnarray}
  \Gamma_{tot}(m) = \Gamma_0\frac{q}{q_0}\frac{m_0}{m}\frac{q^2}{D_1(q,R)}\frac{D_1(q_0,R)}{q_0^2} ,
 \end{eqnarray}
 where $\Gamma_0$ is the width of the $\Xi(1530)$, and $q_0 = q(m_0)$.
 For the fits to the data of Fig.~\ref{fig:XiPimoments}(a), an incoherent background
 function of the form
\begin{eqnarray}
   b = (p\cdot q) \sum_{i=0}^{3} c_i  m^{i}
 \end{eqnarray}
 is included also.

 \begin{figure}[!t]
  \centering\small
  \includegraphics[width=.4\textwidth]{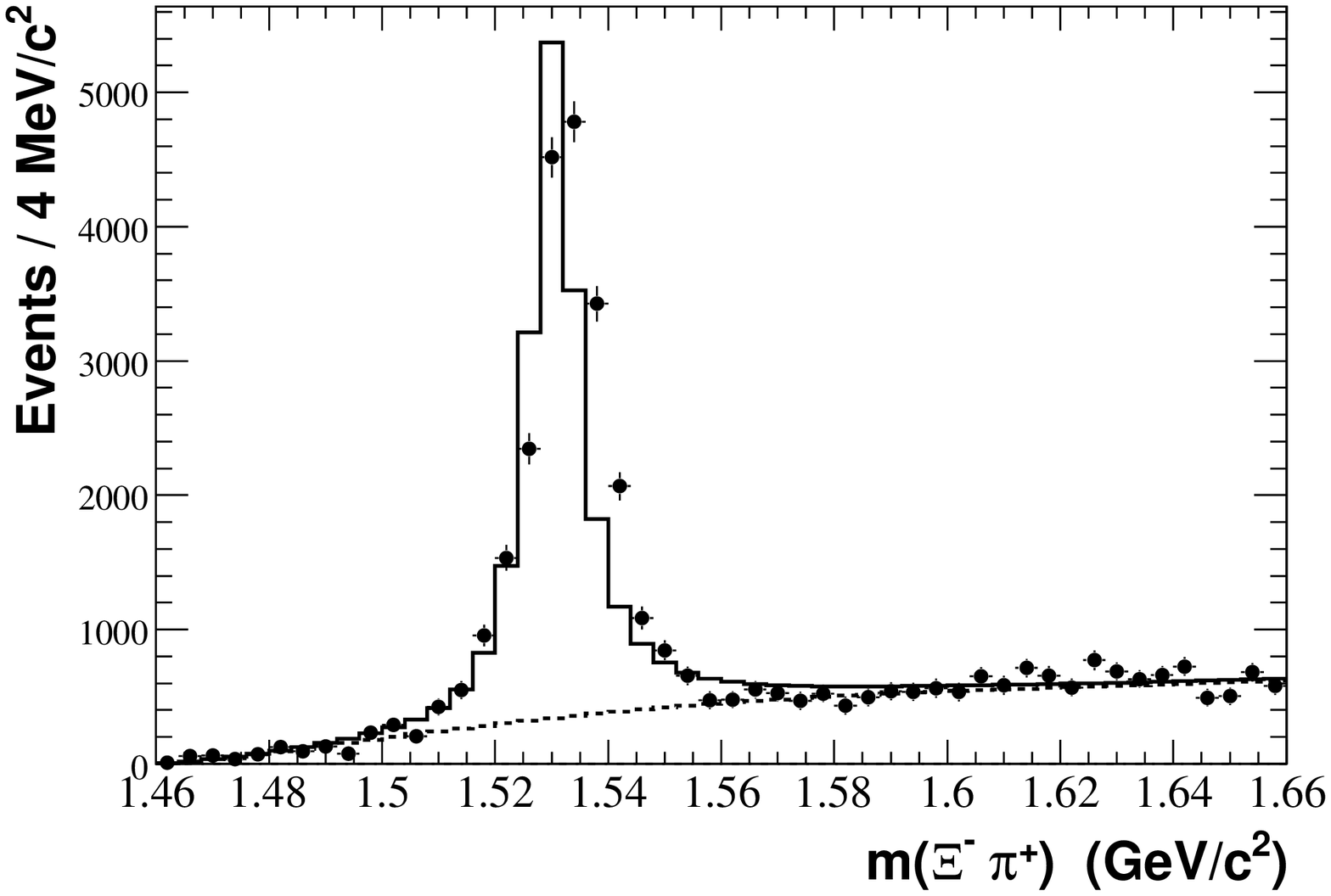}
  \includegraphics[width=.4\textwidth]{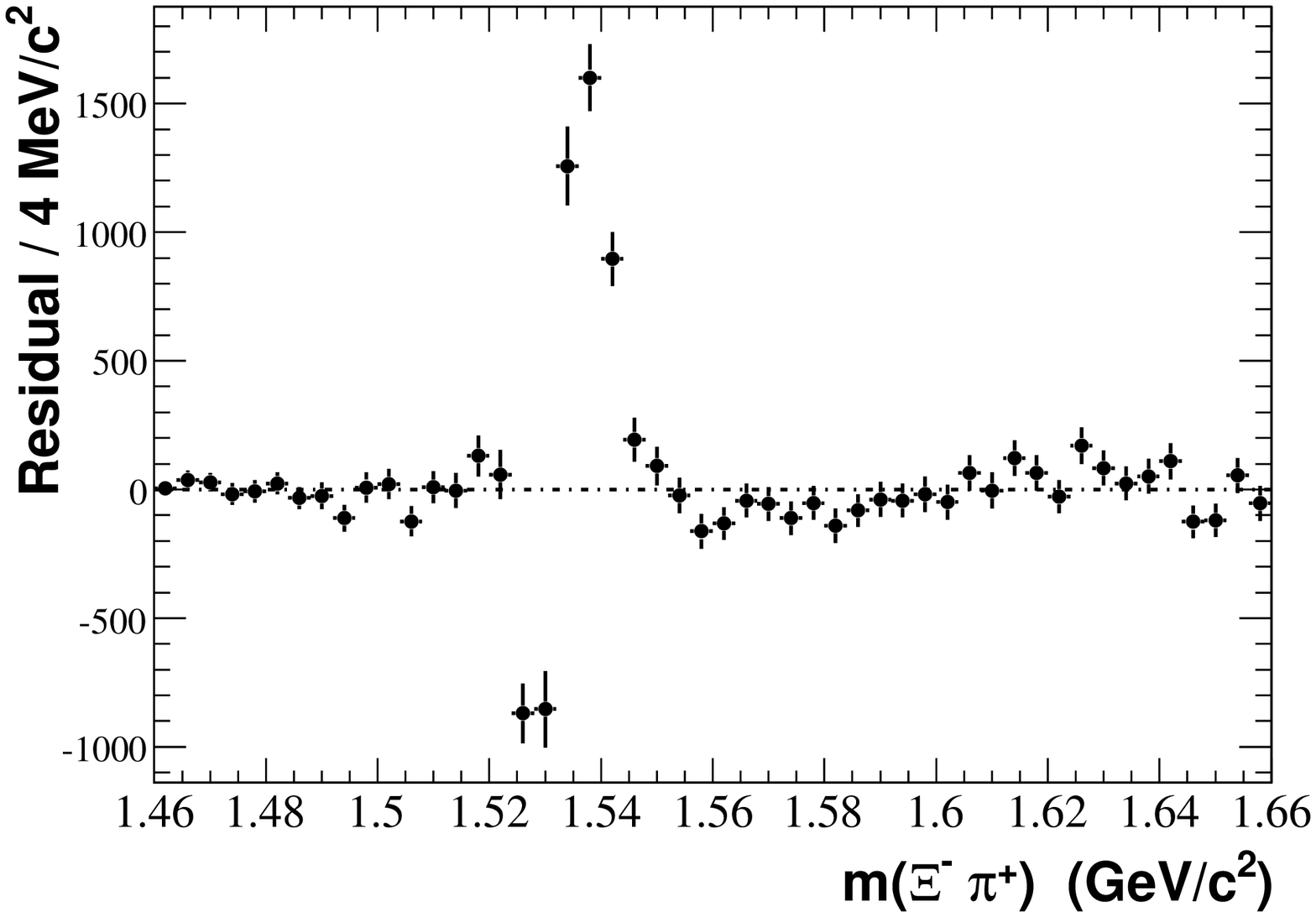}
  \includegraphics[width=.4\textwidth]{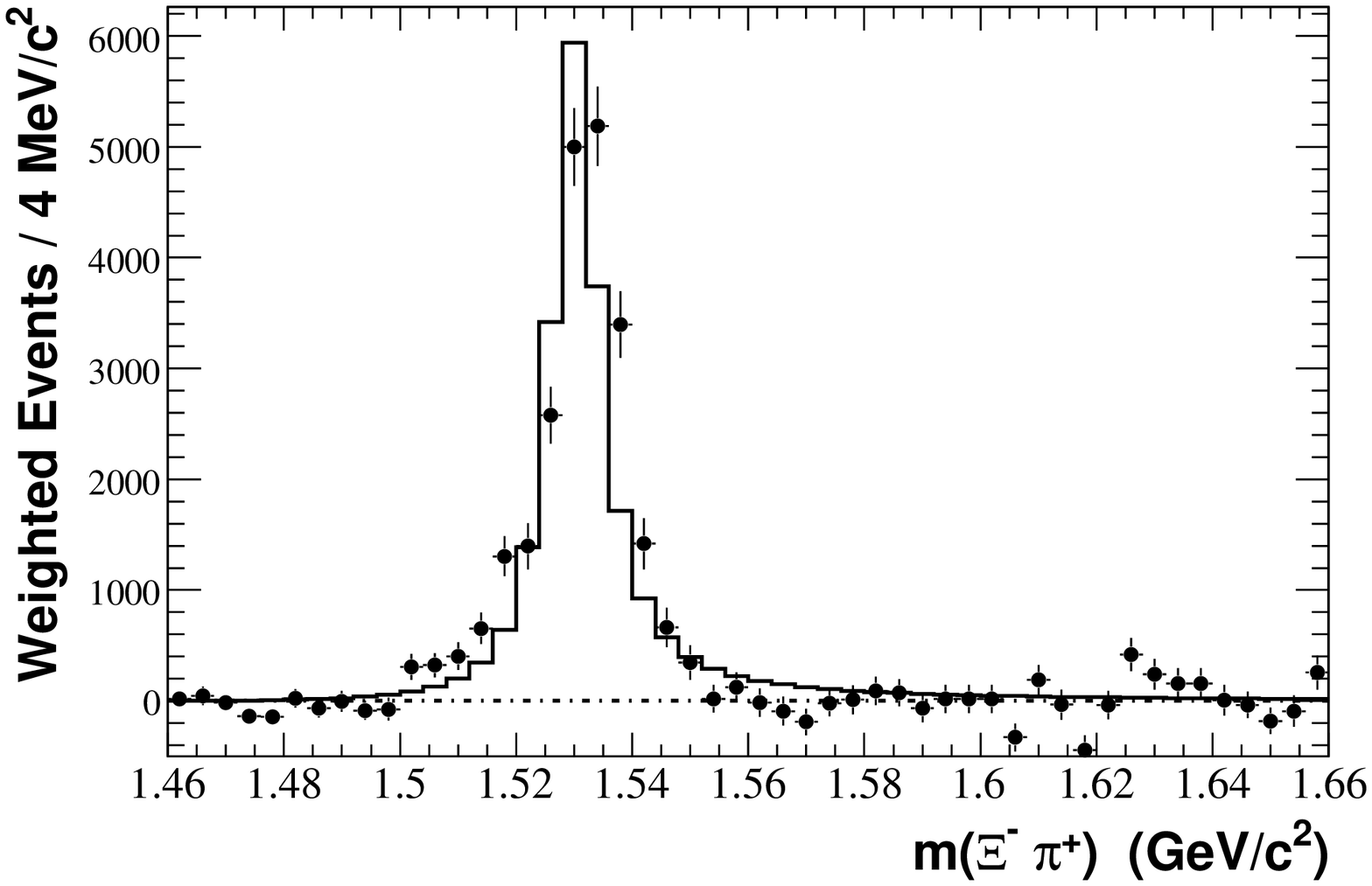}
  \includegraphics[width=.4\textwidth]{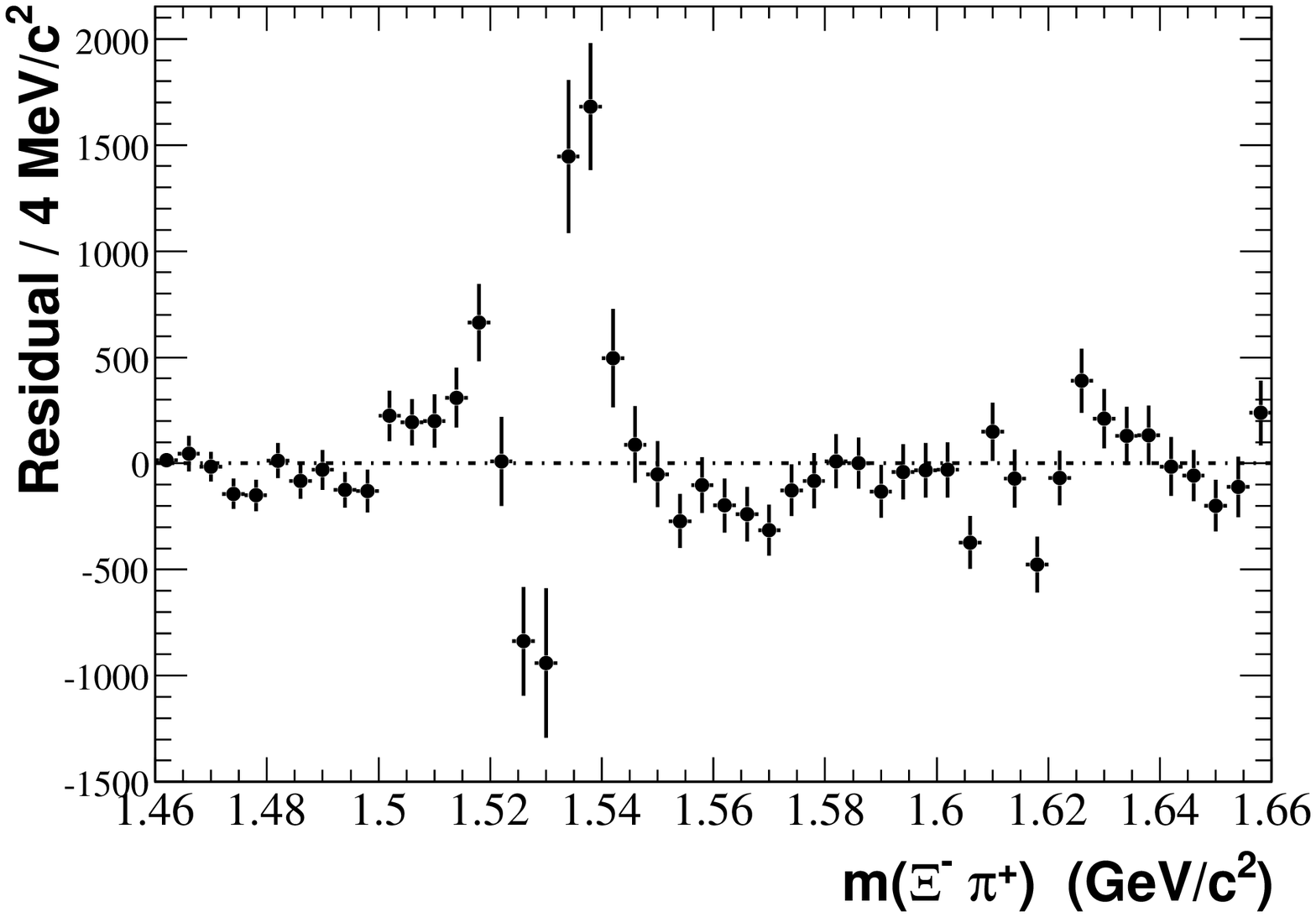}
  \begin{picture}(0.,0.)
   \put(-50,530){\bf{(a)}}
    \put(-50,390){\bf{(b)}}
   \put(-50,250){\bf{(c)}}
    \put(-50,110){\bf{(d)}}
    \end{picture}
 \caption{The efficiency-corrected 
$\Lambda_c^+$-mass-sideband-subtracted (a) $\sqrt{2} P_0(\rm cos\,\theta_{\Xi^-})$ and (c) $\sqrt{10} P_2(\rm cos\,\theta_{\Xi^-})$ 
moment distributions for the $\Xi^{-} \pi^{+}$ system, for the $\Lambda_c^+$ signal region (solid dots).
The fits represented by the histograms are described in the text.  
In (b) [(d)] the difference between the data points and the histogram in (a) [(c)] is shown.}
 \label{fig:P2fix}
\end{figure} 
  In each fit, the fit function is convolved with a mass resolution
 function consisting of two Gaussian distributions with a common center
 and fixed fractional contributions, but with r.m.s. deviation values
 which depend on $\Xi^- \pi^+$ invariant mass. For the resolution function, the resulting half-width-at-half-maximum 
 increases from $\sim 0.5$ MeV/c$^2$ just above threshold, to $\sim 1.5$ MeV/c$^2$ at
 the $\Xi(1530)$, reaching $\sim 2.5$ MeV/c$^2$ at $\sim 1.6$ GeV/c$^2$, so that the
 resolution in the signal region is excellent. 
The degradation of the mass resolution with increasing $\Xi^-\pi^+$ mass should, if anything, cause the 
observed $\Xi(1530)^0$ lineshape to be slightly skewed toward high mass; it follows that this cannot be 
the source of the skewing of the lineshapes of Figs.~\ref{fig:XiPimoments}(a) and~\ref{fig:XiPimoments}(b) toward low mass.  
The convolution procedure takes quantitative account of the resolution behavior, but, 
since the resolution is excellent, this has little impact on the description of the data.
The fit results with
 $\Xi(1530)^0$ mass and width fixed at their PDG values~\cite{PDG2006} are shown
 in Fig.~\ref{fig:P2fix}. In Figs.~\ref{fig:P2fix}(a) and~\ref{fig:P2fix}(c), the dots represent the data 
 of Figs.~\ref{fig:XiPimoments}(a) and~\ref{fig:XiPimoments}(b), respectively, while the histograms represent the 
 mass-resolution-smeared fit functions integrated over the corresponding
 mass intervals. The fit residuals (data $-$ histogram) are shown in Figs.~\ref{fig:P2fix}(b) 
 and~\ref{fig:P2fix}(d), respectively. These show similar very large systematic
 deviations from zero, and the fits have correspondingly poor
 c.l. values ($6\times 10^{-16}$ and $\sim 0$, respectively). With
 the mass and width parameters free in the fits, the c.l. values are
 still poor, and the values obtained differ significantly from the PDG
 values (e.g., $m_0=1534.4\pm 0.1$ MeV/c$^2$ and $\Gamma_0=13.2\pm 0.5$ MeV for
 Fig.~\ref{fig:P2fix}(a)). If the Blatt-Weisskopf radius parameter is allowed to be free,
 an acceptable fit to the mass distribution is not obtained
 (c.l. $\sim 10^{-3}$), the residuals still show systematic deviations from
 zero, and the mass and width values obtained still differ significantly
 from their PDG values. Similarly, the fit to the $P_2$ moment mass dependence
 with mass, width and radius parameters free remains poor (c.l.
 $\sim 10^{-7}$); in addition, the radius parameter increases to
 $\sim 100$ GeV$^{-1}$ (which is equivalent to the use of an $S$-wave Breit-Wigner)
 in an attempt to reproduce the observed lack of skewing
 toward high mass expected for a $P$-wave decay. Since the $P$-wave nature of the decay has been
establised, this is certainly an
 unacceptable result.
                                                                         
  As a check of the signal parametrization of Eq.~16, this
 function (with $R=3$ GeV$^{-1}$) has been used in fits to the published $\Xi^- \pi^+$
 mass distributions from four of the experiments~\cite{london,kirsch,borenstein,baltay} used to obtain
 the PDG mass and width values~\cite{PDG2006}. These are Hydrogen Bubble Chamber
 experiments, and each mass distribution is obtained as the projection
 of the Dalitz plot for the reaction $K^- p \rightarrow \Xi^- \pi^+ K^0$ (Ref.~\cite{kirsch}
 uses some additional contributions). The analysis samples are small
 (125, 350, 324, and 1313 events, respectively), and the details of
 the fit functions used in Refs.~\cite{london,kirsch,borenstein,baltay} are not made clear. 
 It is found that
 the non-resonant background contributions are well-described using only
 the $(p\cdot q)$ phase space factor of Eq.~18, and Eq.~16, convolved with
 a Gaussian of r.m.s. deviation specified for each experiment, is used to represent the
 $\Xi(1530)$ signal. Good c.l. values are obtained, and the resulting weighted
 average mass ($1532.2\pm 0.2$ MeV/c$^2$) and width ($9.9\pm 0.5$ MeV) values
 agree well with those from the PDG~\cite{PDG2006}. This indicates that
 the choice of signal and background functions is not the reason for
 the poor-quality fits to the data of Fig.~\ref{fig:P2fix}. Furthermore, since no
 significant improvement in fit quality is observed in going from
 the $P_2$ fit to the $P_0$ fit, it follows that the difficulty
 in fitting the former cannot be attributed to the presence of an
 $S^{1/2}-D^{5/2}$ interference contribution, since this would not be
 present for the latter.
 
  This striking failure to describe the most obvious feature of the
 $\Xi^- \pi^+ K^+$ Dalitz plot leads to the conclusion that a satisfactory description of the observed
 ($m(\Xi^- \pi^+$), cos$\,\theta_{\Xi^-}$) distribution cannot be obtained in terms of
 amplitudes pertaining solely to the $\Xi^- \pi^+$ system.  The
 difficulties probably result from overlap and interference effects involving amplitudes associated with the 
$K^+\pi^+$ and/or the $\Xi^- K^+$ systems (if the possibility of direct three-body decay is ignored).  
The $K^+\pi^+$ system has $I=3/2$, and has been observed to have only an $S$-wave amplitude, which varies slowly with mass 
in the relevant region ($\leq 1$ GeV/c$^2$).  It seems unlikely that such an amplitude could lead to significant 
distortion of the mass and $\rm cos\,\theta_{\Xi^-}$ dependences of the $\Xi^-\pi^+$ system, 
but the relevant quantitative analysis has not yet been attempted.    
In contrast, the $\Xi^- K^+$ system could have contributions from high-mass $\Lambda$ or $\Sigma^0$ 
resonant structures in the region of overlap with the $\Xi(1530)$ ($>2$ GeV/c$^2$, cf.~Fig.~\ref{fig:Dalitz}(a)).  
Very little is known about such states or their couplings to $\Xi^-K^+$~\cite{PDG2006}, and 
there is no clear evidence for their presence in the Dalitz plots of Fig.~\ref{fig:Dalitz}.  
Indeed the only ``evidence'' for such contributions is the failure of the description of the 
$\Xi(1530)$ region solely in terms of $\Xi^- \pi^+$ amplitudes in the present analysis.  
This seeming impasse might benefit from analyses of related $\Lambda_c^+$ decay processes, such as 
$\Lambda_c^+\rightarrow (\Lambda \eta)\pi^+$ or $\Lambda_c^+\rightarrow (\Lambda \pi^0)\pi^+$, 
although these may suffer from their own particular complications related to other quasi-two-body modes.  

\section{CONCLUSION}

In conclusion, the analysis of the Legendre Polynomial moments of the $\Xi^- \pi^+$ system which 
result from data on the decay $\Lambda_c^+\rightarrow \Xi^- \pi^+ K^+$
has established quite clearly, on the basis of Figs.~\ref{fig:XiPimoments} (b) and~\ref{fig:XiPimoments} (c), 
that the $\Xi(1530)$ hyperon resonance has spin 3/2.  In conjunction with previous 
analyses~\cite{ref:schlein, ref:button},
this also definitively establishes positive parity, and hence that the $\Xi(1530)$ is a $P^{3/2}$ resonance.
However, comparison of the $P_2(\rm cos\,\theta_{\Xi^-})$ moment to the $\Xi^- \pi^+$ mass distribution,
and fits to the angular decay distribution
in the $\Xi(1530)$ region,
indicate that it is necessary to include other $\Xi^- \pi^+$ amplitudes in order to obtain a
complete description of the data.  In particular, the 
observation of a $P_1(\rm cos\,\theta_{\Xi^-})$ moment exhibiting oscillatory behavior in the $\Xi(1530)^0$
region indicates the need for an $S^{1/2}$ amplitude, while providing first evidence for the
expected rapid BW phase motion of the $P^{3/2}$ $\Xi(1530)^0$ amplitude.  
However, a simple model incorporating only these amplitudes and a $D^{5/2}$ amplitude is 
ruled out because of the failure to describe the 
$\Xi(1530)^0$ lineshape. The presence of the $S^{1/2}$ amplitude at high mass,
 and the behavior of the mass distribution at $\sim 1.7$ GeV/c$^2$, suggest
 that a resonant $\Xi(1690)^0$ amplitude may be adding
 coherently to this amplitude, thus leading to the inference of
 spin-parity $1/2^{-}$ for the $\Xi(1690)$.  
It appears that a quantitative description of the $\Xi(1530)$ lineshape, 
and indeed of the entire Dalitz plot, must incorporate these features together 
with amplitude contributions associated with the $K^+ \pi^+$ and/or 
the $\Xi^- K^+$ systems.  An analysis of this complexity will be performed
when the full \babar\ data
sample (integrated luminosity approximately 500 fb$^{-1}$) is available.

\section{ACKNOWLEDGEMENTS}
We are grateful for the 
extraordinary contributions of our \pep2\ colleagues in
achieving the excellent luminosity and machine conditions
that have made this work possible.
The success of this project also relies critically on the 
expertise and dedication of the computing organizations that 
support \babar.
The collaborating institutions wish to thank 
SLAC for its support and the kind hospitality extended to them. 
This work is supported by the
US Department of Energy
and National Science Foundation, the
Natural Sciences and Engineering Research Council (Canada),
the Commissariat \`a l'Energie Atomique and
Institut National de Physique Nucl\'eaire et de Physique des Particules
(France), the
Bundesministerium f\"ur Bildung und Forschung and
Deutsche Forschungsgemeinschaft
(Germany), the
Istituto Nazionale di Fisica Nucleare (Italy),
the Foundation for Fundamental Research on Matter (The Netherlands),
the Research Council of Norway, the
Ministry of Education and Science of the Russian Federation, 
Ministerio de Educaci\'on y Ciencia (Spain), and the
Science and Technology Facilities Council (United Kingdom).
Individuals have received support from 
the Marie-Curie IEF program (European Union) and
the A. P. Sloan Foundation.

{}


\begin{thebibliography} {19}
\bibitem{cc}
{The use of charge conjugate reactions is implied throughout this paper.}
\bibitem{eff}
{Each selected event is weighted by the inverse value of its calculated efficiency, which is parametrized 
as a function of cos$(\theta_{\Xi^-})$ and $m(\Xi^- \pi^+)$ using $\Lambda_c^+\rightarrow \Xi^-\pi^+ K^+$ signal Monte Carlo events.}
\end{thebibliography}
\end{document}